\documentclass[12pt, preprint, trackchanges]{aastex61}
\usepackage{epsfig} %https://www.overleaf.com/project/5b8c792043cfc767092b6c98
\usepackage{amsmath}
\usepackage{multirow}
\usepackage[section]{placeins}
\usepackage[normalem]{ulem}
\usepackage{enumitem}

\newcommand{\ms}{\mbox{m s$^{-1}$}}
\newcommand{\cms}{\mbox{cm s$^{-1}$}}
\newcommand{\kms}{\mbox{km s$^{-1}$}}

\newcommand{\snr}{$\rm S/N$}
\newcommand{\otwo}{$\rm O_2$}
\newcommand{\ntwo}{$\rm N_2$}
\newcommand{\cotwo}{$\rm CO_2$}
\newcommand{\chfour}{$\rm CH_4$}
\newcommand{\nox}{$\rm NO_X$}

\newcommand{\angstrom}{\textup{\AA}}

\received{}
\revised{}
\accepted{}
\submitjournal{ApJ}

\shorttitle{Tellurics}
\shortauthors{Leet}

\begin{document}

\title{Towards a Self-calibrating, Empirical, Light-Weight Model for Tellurics in High-Resolution Spectra}
\author[0000-0001-5847-9147]{Christopher Leet}
\affiliation{Yale University, 52 Hillhouse, New Haven, CT 06511, USA}

\author[0000-0003-2221-0861]{Debra A. Fischer}
\affiliation{Yale University, 52 Hillhouse, New Haven, CT 
06511, USA}

\author[0000-0002-9916-8393]{Jeff A. Valenti}
\affiliation{Space Telescope Science Institute, 3700 San Martin Dr., Baltimore, MD 21218 USA}

\correspondingauthor{Christopher Leet}
\email{christopher.leet@yale.edu}


\begin{abstract} 
{To discover Earth analogs around other stars, next generation spectrographs must measure radial velocity (RV) with 10 cm/s precision.} {To achieve 10cm/s precision, however, the effects of telluric contamination must be accounted for.} {The standard approaches to telluric removal are: (a) observing a standard star and (b) using a radiative transfer code. Observing standard stars, however, takes valuable observing time away from science targets. Radiative transfer codes, meanwhile, rely on imprecise line data in the HITRAN database (typical line position uncertainties range from a few to several hundred m/s) and require difficult-to-obtain measurements of water vapor column density for best performance. To address these issues, we present SELENITE: a SELf-calibrating, Empricial, Light-Weight liNear regressIon TElluric model for high-resolution spectra. The model exploits two simple observations: (a) water tellurics grow proportionally to precipitable water vapor and therefore proportionally to each other and (b) non-water tellurics grow proportionally to airmass. Water tellurics can be identified by looking for pixels whose growth correlates with a known calibration water telluric and modelled by regression against it, and likewise non-water tellurics with airmass. The model doesn't require line data, water vapor measurements and additional observations (beyond one-time calibration observations), achieves fits with a $\chi^2_{red}$ of 1.17 on B stars and 2.95 on K dwarfs, and leaves residuals of $1\%$ (B stars) and $1.1\%$ (K dwarfs) of continuum. Fitting takes seconds on laptop PCs: SELENITE is light-weight enough to guide observing runs.}

%Telluric contamination of optical spectra {is an important source of} error for the next generation of extremely precise radial velocity programs.  Here, we present an empirical method for identifying and modeling telluric features in the optical spectrum (400 - 680 nm) based on {the following observations}: the line depth of water telluric features in a given spectrum is dominated by the changing column density of precipitable water vapor; the growth of any unsaturated water telluric feature is directly proportional to the growth of other water tellurics; and the growth of non-water telluric lines is linearly correlated with airmass. {Regression analysis against an ensemble of pre-identified water lines allows us to identify and model water lines and regression analysis against airmass allows us to model non-water telluric lines.} A set of B star spectra observed with the CHIRON spectrograph at different airmass and on nights with variable amounts of precipitable water vapor was used to derive the linear coefficients needed to generate a complete telluric model in the optical bandpass. The accuracy of this empirical model depends on the quality of spectra in the training set. Here we show that for \snr\ $\sim 100$, tellurics as shallow as 2\% can be identified.   
 \end{abstract}

\section{Introduction}
To expand the success of exoplanet searches, next generation spectrographs are aiming for sub-meter-per-second precision in radial velocity measurements. If the 10 \cms\ instrumental precision goal of the Echelle SPectrograph for Rocky Exoplanets Search and Stable Spectroscopic Observations \citep[ESPRESSO][]{Pepe2013} and the EXtreme PREcision Spectrograph \citep[EXPRES][]{Jurgenson2016} is reached, we will be able to detect small rocky planets orbiting in the habitable zones of their host stars. Such high precision requires extraordinary new fidelity in spectroscopic data: high resolution, high \snr{} and greater instrumental stability. In addition to controlling instrumental errors, success requires accounting for any systematic temporal changes in the spectral line profiles, which can arise from photospheric velocities or telluric contamination \citep{Fischer2016}.

Most work on modeling telluric contamination has been tested at near infrared wavelengths where the telluric line depths are comparable to stellar absorption lines. However, the next generation optical spectrographs aiming for 10 \cms\ radial velocity precision will be affected by time variable microtellurics that raster across the stellar spectrum {because of} barycentric velocity shifts. If we do not identify pixels that are producing the small perturbations to spectral line profiles, then microtellurics may  dominate the error budget for extreme precision radial velocity programs. 

\section{Telluric Spectra}
Atomic and molecular species in the Earth's atmosphere interact with solar radiation and produce absorption and emission lines that are imprinted in stellar spectra obtained with ground-based spectrographs. The non-water constituents (e.g., \ntwo, \otwo, Ar, Ne, He) are well-mixed, and maintain a nearly fixed element ratio throughout the troposphere, stratosphere, and mesosphere. The concentration of some non-water species (\cotwo, \chfour, \nox) exhibit seasonal changes or modulation from post-industrial human activities. However, these gases have stable concentrations on timescales of (at least) several days. In contrast, 99\% of atmospheric water vapor is confined to the troposphere and exhibits both temporal and spatial variability that can change by more than 10\% on timescale of an hour \citep{BlakeShaw2011}. 

\begin{figure}
\epsscale{1.1} 
	\plottwo{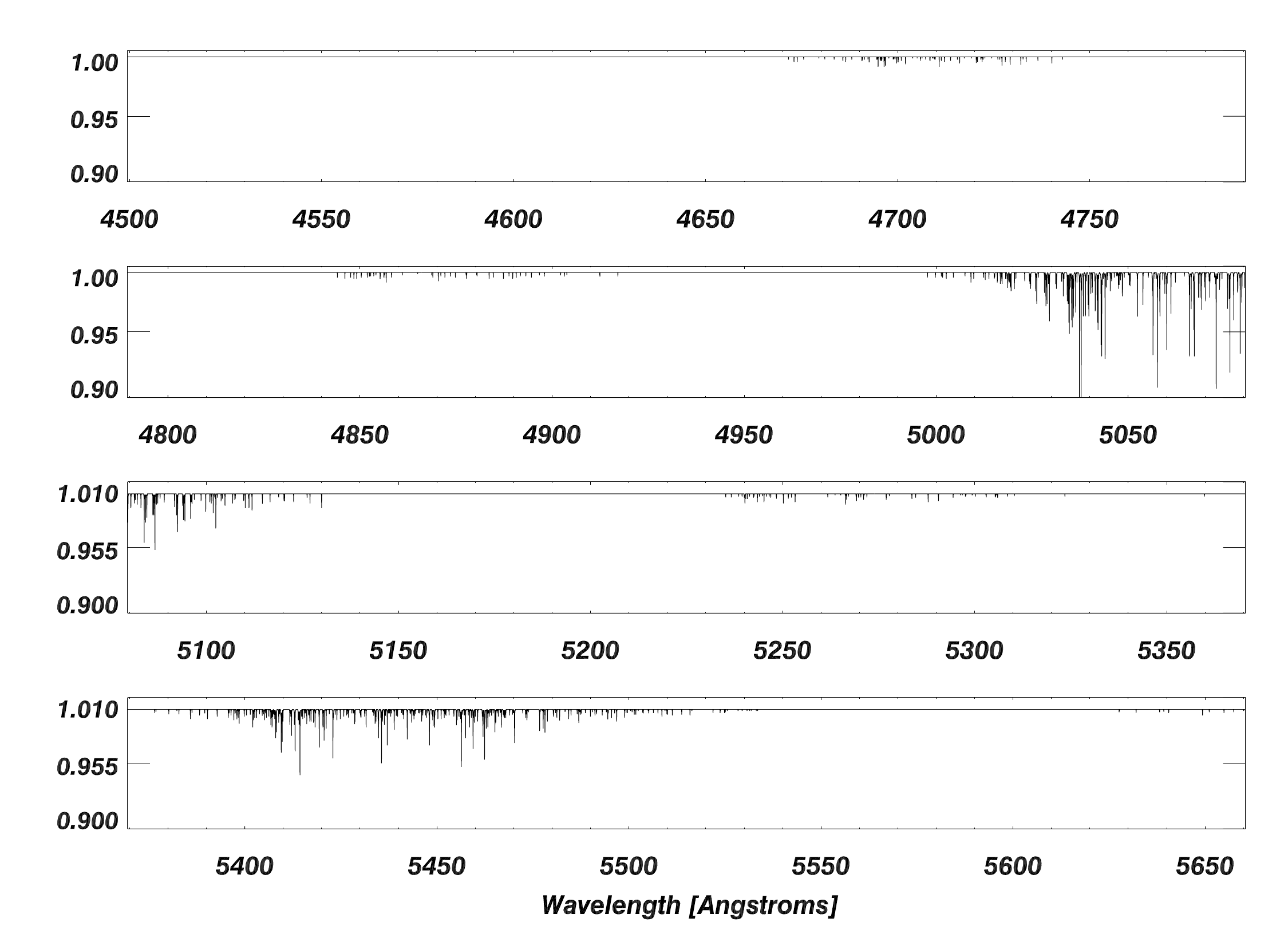}{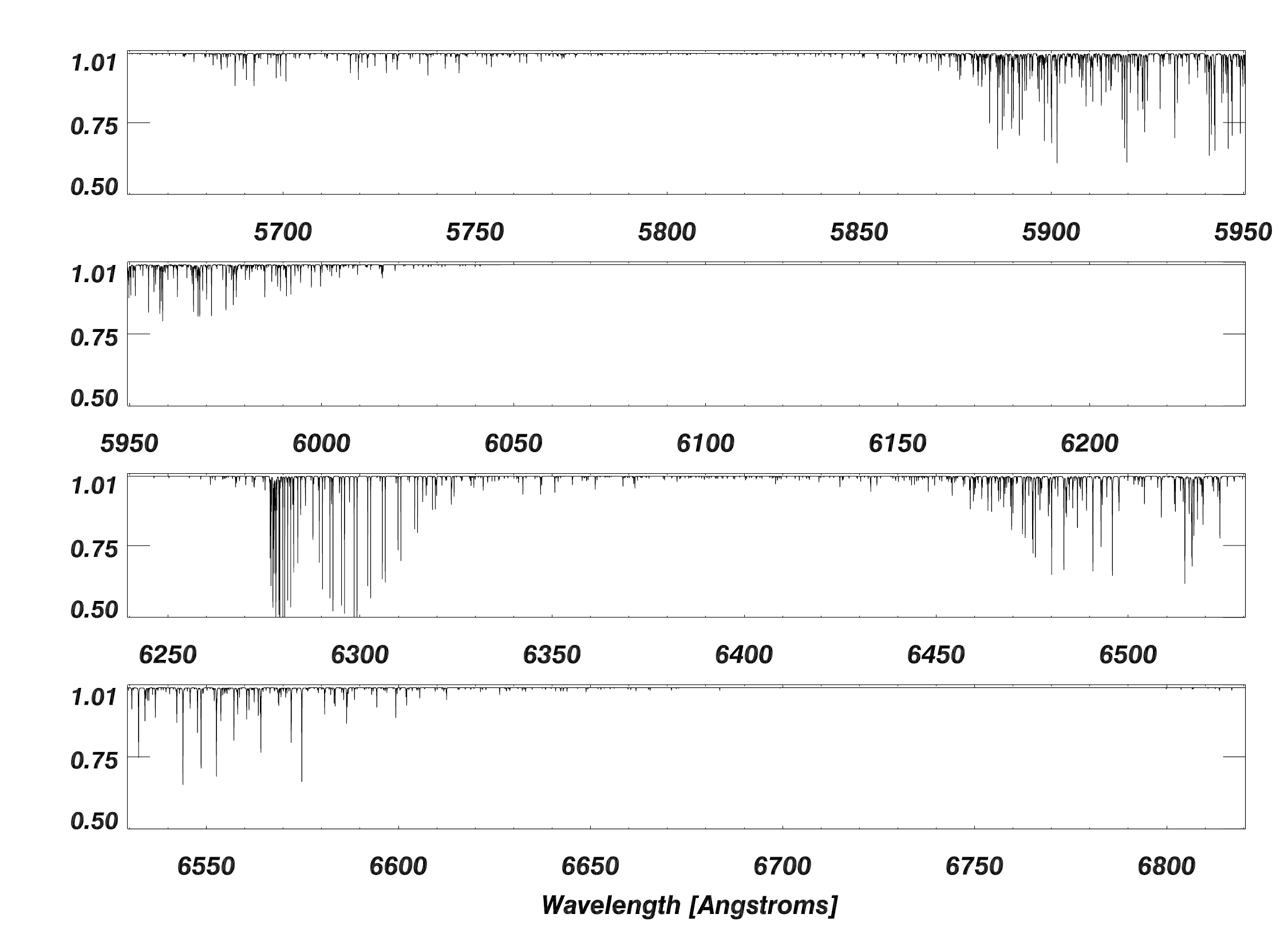}
	\caption{The FTS solar spectrum from 4500 - 6800\angstrom{} has a resolution, $R\sim 600,000$ and \snr $> 1000$. This high fidelity spectrum shows shallow microtelluric atmospheric lines (note the difference in the intensity range for the left and right panels). About 20\% of pixels in the optical spectrum have lines deeper than 0.1\% and 4\% of the pixels have microtellurics deeper than 1\%.
	\label{fig:nso_microtell}}
\end{figure}

Figure \ref{fig:nso_microtell} shows the telluric spectrum with a wavelength range of $4500 - 6800\angstrom{}$ obtained with the Fourier Transform Spectrograph (FTS) from the National Optical Astronomical Observatory \citep[NOAO][]{Wallace1993}. The strongest telluric lines are found redward of about 6800\AA\ and present a particular challenge for radial velocity measurements in the near infrared. However, he high \snr\ and resolution of the FTS telluric spectrum shows that the optical spectrum is peppered with microtelluric lines with depths that are only a few percent of the continuum. Many of the lines shallower than 1\% in Figure~\ref{fig:nso_microtell} will disappear when convolved with the instrumental line spread function (LSF) of high-resolution ($R~100,000$) echelle spectrographs. The surviving microtelluric lines are, however, very difficult to discern when superimposed on stellar spectra. Even for stars with constant radial velocity, the barycentric velocity of the Earth causes the telluric lines to raster across stellar absorption lines with annual amplitudes up to 30 \kms, producing small, but systematic, time-variable line profile variations. Optical RV programs aiming for 10 - 20 \cms\ precision will need to account for microtelluric lines because they introduce errors that exceed the target RV precision \citet{Cunha2014}.

% 20% of pixels have lines deeper than 0.1%
% 4% of pixels have lines deeper than 1%
% 2% of pixels have lines deeper than 2%

\section{Current Best Practices}
{Since telluric contamination is a serious error source for high precision spectroscopy, there is a rich literature of practices for telluric modelling. These practices fall into three categories: \textbf{(a)} modelling using telluric standard stars (Section~\ref{subsec:telluric_standard_stars}), \textbf{(b)} modelling using radiative transfer codes (Section~\ref{subsec:radiative_transfer_codes}) and \textbf{(c)} modelling using principle component analysis (Section~\ref{subsec:pca}). Finally, we discuss the literature surrounding a new challenge in telluric modelling: microtelluric modelling (Section~\ref{subsec:microtellurics_challenge}).}

%One approach for minimizing the impact of telluric contamination on radial velocity measurements is to simply down-weight or mask out pixels that are affected by known telluric lines. This results in a loss of many pixels in the near infrared spectrum.

\subsection{Telluric Standard Stars}
\label{subsec:telluric_standard_stars}
{The classical approach to removing telluric absorption features is to observe a telluric standard star close in time and airmass to the science object \citep{Vacca2003,1986:Vidal}. The science target's spectrum is then divided by the spectrum of the standard star. Typically, early type stars from early A to late B are chosen as standard stars because they exhibit few and weak metal lines, and their rapid rotation helps smear out the lines that remain. The high \snr\ afforded by bright stars means that with high spectral resolution, even shallow telluric lines are discernible. These stars have the drawback that their strong hydrogen absorption features at the Brackett and Paschen lines blend with their tellurics \citep{Rudolf2016}. As an alternative, a solar type star can be used as a telluric standard using a high-resolution solar spectra \citep{Maiolino1996}.}

{Using any standard star as a telluric reference model has several well known drawbacks. \textbf{First,} it takes away precious observing time from an observation's science targets, especially when high \snr\ requirements are to be met \citep{Seifahrt2010}. \textbf{Second,}  its accuracy is limited by how well the standard star's spectrum is known. Early type stars often display spectral features such as oxygen or carbon lines in the near-infrared. Similarly, absorption line depths of solar-type stars may deviate from the solar FTS atlas due to metal abundance or surface temperature deviations, leaving residuals from the star's intrinsic features in the telluric model \citep{Rudolf2016}. Compounding this problem, the need to pick a star close to the science target often forces the observation of less well known stars. \textbf{Finally,} for telescopes with an adapative optics system (e.g. CRIRES), the change in source brightness between the science target and the standard star will affect the instrumental profile \citep{Seifahrt2010}. In practice, \cite{UlmerMoll2018} find that standard stars consistently underperform other telluric removal approaches.}

%\deleted{When telluric lines are well resolved, another approach is to divide by a telluric standard star \citep{Vacca2003}. However, it is the product of the stellar flux and the telluric absorption that is convolved with the instrumental line spread function, so this approach is not mathematically rigorous and the precision of this operation is probably not adequate for precise radial velocity work. Telluric standard stars are typically bright, hot, and rapidly rotating stars (spectral types OBA). Hot stars have nearly featureless spectra and rapid rotation helps to smear out any stellar absorption lines. The high \snr\ afforded by bright stars means that with high spectral resolution, even shallow telluric lines are discernible as sharp features. The telluric star should be located at a small angular distance from the program star so that the atmospheric column is similar. However, telluric standards often have strong emission reversals in the cores of Hydrogen ${\rm H\alpha}$ and the sodium D1 and D2 lines making these wavelength regions unsuitable. This approach can be relatively expensive since telescope time must be devoted to observing telluric standards throughout the night, instead of observing program stars.}

\subsection{Radiative Transfer Codes}
\label{subsec:radiative_transfer_codes}
Another approach is to use line-by-line radiative transfer model (LBLRTM) codes to model telluric lines. This technique requires accurate atmospheric temperature and pressure profiles, an excellent model for the spectrograph line spread function, and a complete and accurate atomic line data base. The atmospheric inputs to these codes have benefited from the commercial interest and investments in making more accurate weather predictions. Most radiative transfer codes, including TERRASPEC \citep{Bender2012}, Transmissions Atmosph{\'e}riques
Personnalis{\'e}es Pour l{'}AStronomie \citep[TAPAS][]{Bertaux2014}, Telfit \citep{Gullikson2014} and Molecfit \citep{Smette2015} use the HIgh Resolution TRANsmission line database \citep[HITRAN][]{Rothman2013} and are able to model non-water telluric lines with an accuracy of {around 2\%}. %{\cite{Seifahrt2010} also find that most radiative transfer codes achieve an accuracy of 2\%.}

{Unfortunately, radiative transfer codes also suffer from documented drawbacks. \textbf{First}, radiative transfer codes are limited by imprecise line data in the HITRAN database. The uncertainty in each HITRAN line's position is typically a few to several hundred m/s, but can be up to multiple km/s. HITRAN line strengths are rarely accurate to the 1\% level \citep{Seifahrt2010}. \cite{Rudolf2016} also remark on this problem when modelling tellurics in the near IR.}

{\textbf{Second}, radiative transfer codes often struggle to model water lines.} \citet{Bertaux2014} identify some cases in TAPAS where two adjacent water lines required different amounts of water for an adequate model. This is clearly non-physical (there is only one column density of water), but the authors are uncertain why this discrepancy appears. {\cite{Rudolf2016} note that HITRAN has imperfect water line information and induce substantial residuals in their radiative transfer code.} 

\subsection{Principal Component Analysis}
\label{subsec:pca}
\citet{Artigau2014} investigated the use of principal component analysis (PCA) for empirically modeling telluric lines at near infrared wavelengths. They used observations of hot, rapidly rotating stars to build a library of telluric standards with a range of water column density and air mass. The first five principal components of the telluric absorption features were used to fit telluric lines in spectra of program stars using least squares fitting. This empirical approach self-calibrates spectra and avoids the need for atomic line data or estimates of water column density. {We believe that PCA's empirical approach is on the right track. However, PCA is a very generic model, and could benefit by incorporating the well-studied physics of telluric line formation.} {By introducing principled physical priors, we aim to improve the sophistication of this approach.}

% \deleted{\cite{Artigau2014} tested their approach on 1526 observations of Tau Ceti that were averaged to 73 epochs. When they used only the near infrared portion of the spectrum, the radial velocity RMS improved from 10 \ms\ to 1.6 \ms\ after PCA-based modeling. Using the entire optical spectrum with pixels selected by the HARPS %\citep[High Accuracy Radial velocity Planetary Searcher;][]{Mayor2003} 
%cross-correlation mask, which is designed to eliminate most telluric features, the RMS decrease was more modest and dropped by 14 \cms\ from 1.07 \ms\ to 0.93 \ms. While the PCA methodology is transferable, the telluric libraries should be reconstructed for each spectrograph because of differences in resolution, spectral format, and line spread functions.} 

\subsection{The Challenge of Microtellurics}
\label{subsec:microtellurics_challenge}
Most methods for modeling telluric lines have been applied to lines that are redward of 6800\AA. The telluric features at these red wavelengths are easier to identify, both because the telluric lines are deeper and the density of stellar lines is decreasing. Currently there is not a robust method for modeling microtellurics. 
Unfortunately, simulations by \citet{Cunha2014} show that if ignored, microtelluric contamination in the optical spectrum will introduce RV errors between 0.2 - 1.0 \ms, swamping the error budget of next generation RV surveys. \citet{Cunha2014} modeled microtelluric lines in HARPS optical spectra using TAPAS, an online service that simulates atmospheric transmission with input from the ETHER Atmospheric Chemistry Data Centre, atomic line data from HITRAN, and an LBLRTM code \citep{Bertaux2014}. The atmospheric temperature and pressure model for the geographic region near La Silla is updated every six hours, and the model with the closest match in time to observations is adopted with small empirical adjustments to water vapor column density. Based on simulations with synthetic spectra, \citet{Cunha2014} expected that the improvement in RV precision for most stars would be in the range of 10 - 20 \cms. {Achieving RV  accuracies of 10 \cms{} necessitates accurate modelling of microtellurics.}

%The improvement with this approach was not possible to demonstrate in practice, because the errors from microtellurics are generally smaller than the 0.5 - 1.0 \ms\ precision at HARPS.  

\section{SELENITE: A Self-Calibrating Linear Regression Model}
{We now describe SELENITE's telluric model. Since water and non-water tellurics exhibit different behavior \citep{Hadrava2006}, SELENITE} {treats their lines} {separately, and so we develop the model as follows. First, we describe the training data used to illustrate and evaluate SELENITE (Section~\ref{subsec:training_data}). We proceed to describe the model for water tellurics (Section~\ref{subsec:water_tellurics}) and evaluate its performance on the B star HR3982 (Section~\ref{subsec:water_results}). We then describe the model for non-water tellurics (Section~\ref{subsec:non_water_tellurics}) and evaluate its performance (Section~\ref{subsec:non_water_results}), before finally combining the two halves and applying them to Alpha Centaur B, a K dwarf with significant stellar features (Section~\ref{subsec:K_dwarf_modelling}).}

%Important differences in the formation of water and non-water lines that motivated us to describe our modeling process for water and non-water telluric lines separately. In Section~\ref{sec:water_tellurics} we describe a model for water tellurics that requires pre-identification of a pixel where the fractional contamination by water tellurics can be measured. In Section~\ref{sec:non-water}, we describe a linear regression against airmass that is effective for identifying non-water lines. Ultimately, the method that we have developed accounts for both water and non-water tellurics simultaneously. {Note that this method excludes treatment of saturated lines: in our calibration data, taken in the optical (4000\angstrom{}-6800\angstrom{}) at airmasses up to 1.8, all lines were weaker than 50\% of the continuum.}

\subsection{Training Data}
\label{subsec:training_data}
The training data included 51 spectra of rapidly rotating B stars observed with the fiber-fed CHIRON spectrograph \citep{Tokovinin2013}, which is located at 1.5-m telescope at the Cerro Tololo Interamerican Observatory (CTIO). The B-type stars are ideal for this calibration because they are bright and have few spectral lines, providing high \snr\ spectra that are relatively easy to continuum normalize. The iodine cell that is used for Doppler measurements with CHIRON was not in the light path for any of these observations. These spectra were obtained with the narrow slit mask, which yields a spectral resolution, $\lambda / \delta\lambda$ of ${\rm R = 140,000}$ and exposure times were set to reach a typical \snr\ of 100. The air mass for each observation was recorded in the FITS header; however, no information was available regarding the PWV or other atmospheric conditions.

\citet{Figueira2010} demonstrate long-term stability of telluric lines at the level of 10 \ms\ (corresponding to 0.01 of a pixel) at the La Silla Observatory using the environmentally stabilized and fiber-fed HARPS spectrograph. The CHIRON spectrograph does not have the stability of HARPS, and the spectral format can drift by a fraction of a pixel from night-to-night. To correct for these small drifts, the spectral orders were cross-correlated to align the telluric absorption lines.

\subsection{Water Tellurics}
\label{subsec:water_tellurics}

\subsubsection{The Theory of Water Tellurics}
\label{ref:water_telluric_theory}
Each water vapor line has a specific absorption coefficient, $\sigma$, which depends on fundamental atomic and molecular line data, including the $\log (gf)$ value, excitation potential, and the partition function. The line strength of water tellurics also changes with the number of absorbers along the line of sight, or the column density. The radiative transfer equation for the intensity of light with wavelength $\lambda$ passing through a plane-parallel atmosphere with a single species of absorber is:

\vspace{-3mm}
\begin{align}\label{eqn:lnI}
I_\lambda = I_{\lambda,0} \, e^{-\sigma_\lambda \cdot n_j \cdot z} \\
\ln {I_\lambda} = -\sigma_\lambda \cdot n_j \cdot z 
\end{align}
\vspace{-3mm}

\noindent
where $I_{\lambda,0}$ and $I_\lambda$ are the initial and final intensity, $\sigma_\lambda$ is the effective cross-section for absorption, and $n$ is the average number density of water vapor absorbers. {Path length, $z$, is measured in units of airmass at zenith. The column density of water vapor, PWV is $n_j \cdot z$. If a spectrum is normalized, ($I_{\lambda, 0} = 1.0$), the natural logarithm of its line intensity is proportional to the average absorption cross-section and the number of absorbers.
While each water line will have a unique absorption cross-section, all water lines in an observation will share the same PWV ($n \cdot z$).The depth of any two water lines is therefore linearly related: \textbf{by measuring the depth of an arbitrary water line (or set of lines), we can predict the depth of every other water line in the spectrum.} We refer to the water telluric used to construct the telluric spectrum as the calibration telluric, and the pixel at the core of the calibration line as the calibration pixel.}

\begin{figure}[!tbhp]
  \includegraphics[trim=0in 7.5in 22.25in 0in, width=3in]{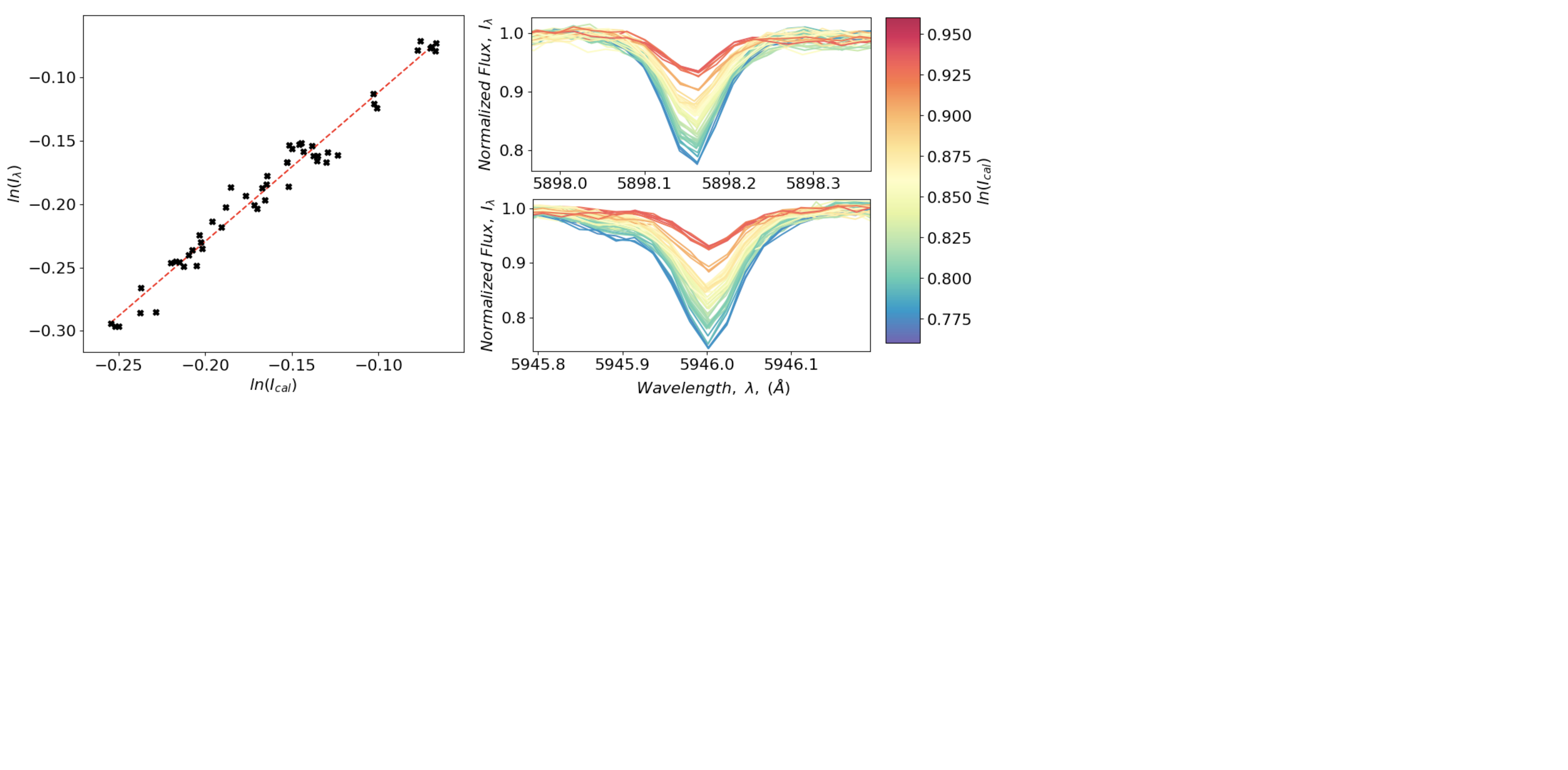}
  \caption{(left) Correlation between $\ln({I_\lambda})$ in a pixel containing a 
  telluric signal at $5946.0\angstrom$ and the central pixel of a known water telluric calibration line, $\ln({I_{cal}})$, at $5898.16\angstrom$. (right-top) The time series of 51 CHIRON spectra for the calibration line (5898.16\angstrom), color-coded by line depth. (right-bottom) The time series of 51 CHIRON spectra at 5946.0\angstrom{}, colored-coded by the calibration line's depth to emphasize the correlated growth of these two lines.}
  \label{fig:o29-o44-growth}
\end{figure}

As an example, Figure~\ref{fig:o29-o44-growth} shows two water telluric lines 
from the set of training spectra. Both sets of spectra (Figure~\ref{fig:o29-o44-growth} right, top and bottom) have been color-coded by the intensity of the pixel at $\lambda = 5898.16$\angstrom{}, emphasizing the correlated line growth. In the left panel of Figure~\ref{fig:o29-o44-growth}, the correlation between the logarithm of the pixel intensity for these two water telluric features is shown to be linear, with a Pearson Correlation Coefficient (PCC) of 0.99, and {the fitted regression line has residuals of 0.0085, comparable to the average deviation of the continuum from unity (0.01)}.

{We now derive the precise relationship between the depths of any two water tellurics.} From the radiative transfer equation, the intensities of any pair of water lines, $(I_{\lambda_i}, I_{\lambda_{cal}})$, grow proportionally to each other in log space. Since the average number density of water absorbers and the airmass is a constant at any time $t_i$, the constant of proportionality between the growth of two lines, as shown in Equation~\ref{eqn:slope_a}, can be physically interpreted as the ratio between the absorption cross-section at two wavelengths: $\sigma_{\lambda_i}/\sigma_{\lambda_{cal}}$. We denote this constant of proportionality as $m^{\lambda_i}_\text{cal}$. 

\begin{equation}
\label{eqn:slope_a}
\frac{\ln{I_{\lambda_{i,t2}}} - \ln{I_{\lambda_{i,t1}}}}{\ln{I_{\lambda_{cal,t2}}} - \ln{I_{\lambda_{cal,t1}}}} = \frac{\sigma_{\lambda_i} \left[n_{t2} \cdot z_{t2} - n_{t1} \cdot z_{t1}\right]}{\sigma_{\lambda_{cal}} \left[n_{t2} \cdot z_{t2} - n_{t1} \cdot z_{t1}\right]} = \frac{\sigma_{\lambda_i}}{\sigma_{\lambda_{cal}}} 
\equiv m^{\lambda_i}_\text{cal}
\end{equation}

A similar linear regression is carried out to empirically relate every other pixel in the spectrum to the calibration pixel, implying an equation of the form {$\ln {I_{\lambda_i}} = m^{\lambda_i}_{\lambda_{cal}} \ln {I_{\lambda_{cal}}} + b$.} During this process, the y-intercept was always found to be zero, simplifying the regression model to:

\begin{equation}\label{eqn:slope_b}
\ln{I_{\lambda_i}} = m^{\lambda_i}_\text{cal} \ln{I_{\lambda_{cal}}}
\end{equation}

{One exception to the above are saturated tellurics, which have left the linear regime of growth and do not obey Equation~\ref{eqn:lnI}. In both our water and non-water analysis, however, we find no telluric deeper than 50\% of continuum between 4500\angstrom{}-6800\angstrom{} and so no saturated telluric. Saturated tellurics are therefore considered outside of this paper's scope. Another exception to the above is variations in the instrumental line spread function (LSF) over time changing a telluric's profile. SELENITE does not model instrumental errors, and these variations can only be handled by observing new training data under the new LSF. Fortunately, at CHIRON's resolution tellurics are marginally resolved, attenuating LSF changes. In practice, CHIRON's LSF is relatively stable over years, allowing 2012 K-dwarf observations to be fit by a model built on 2014 B star observations (Section~\ref{subsec:K_dwarf_modelling}).}

{The correlated growth of water tellurics can also be exploited to identify water tellurics. The PCC of each pixel's growth with the calibration pixel can be measured, and each pixel whose PCC exceeds a threshold, $k$, can be flagged as containing a water telluric. Usefully, SELENITE can discover new water tellurics not contained in HITRAN and correct the position of HITRAN's water tellurics.}

Three additional tests are applied to pixels with PCC $> k$ to eliminate false positives: First, the line spread function for CHIRON has a full width half maximum of 3 pixels. Therefore, we require a minimum of three consecutive pixels with PCC values that exceed $k$. Single or double pixels are assumed to be spurious. Second, because telluric lines have Gaussian profiles, the cluster of flagged pixels must pass a peak detection algorithm. Finally, the high resolution FTS solar spectrum (Figure~\ref{fig:nso_microtell}) indicates that telluric lines appear in clusters rather than as single isolated lines. Any isolated telluric without another telluric within 10\angstrom{} is therefore rejected.

\subsubsection{Establishing a PCC Threshold}\label{sec:bootstrap}
The threshold PCC ($k$) for flagging pixels with a telluric signal must be chosen to minimize both the number of both spurious detections (false positives) and the number of missed telluric lines (false negatives). This critical step ensures that the model telluric spectrum will have the highest possible fidelity. If spurious features are included in a model, they will be used to assign zero weight pixels, resulting in lost data for the radial velocity cross-correlation. If telluric features are missed in a model, they will remain in the stellar spectrum and increase the radial velocity errors. 

\begin{figure}[!tbhp]
  \centering
  \includegraphics[clip, trim=0in 7.25in 9.25in 0in, width=7in]{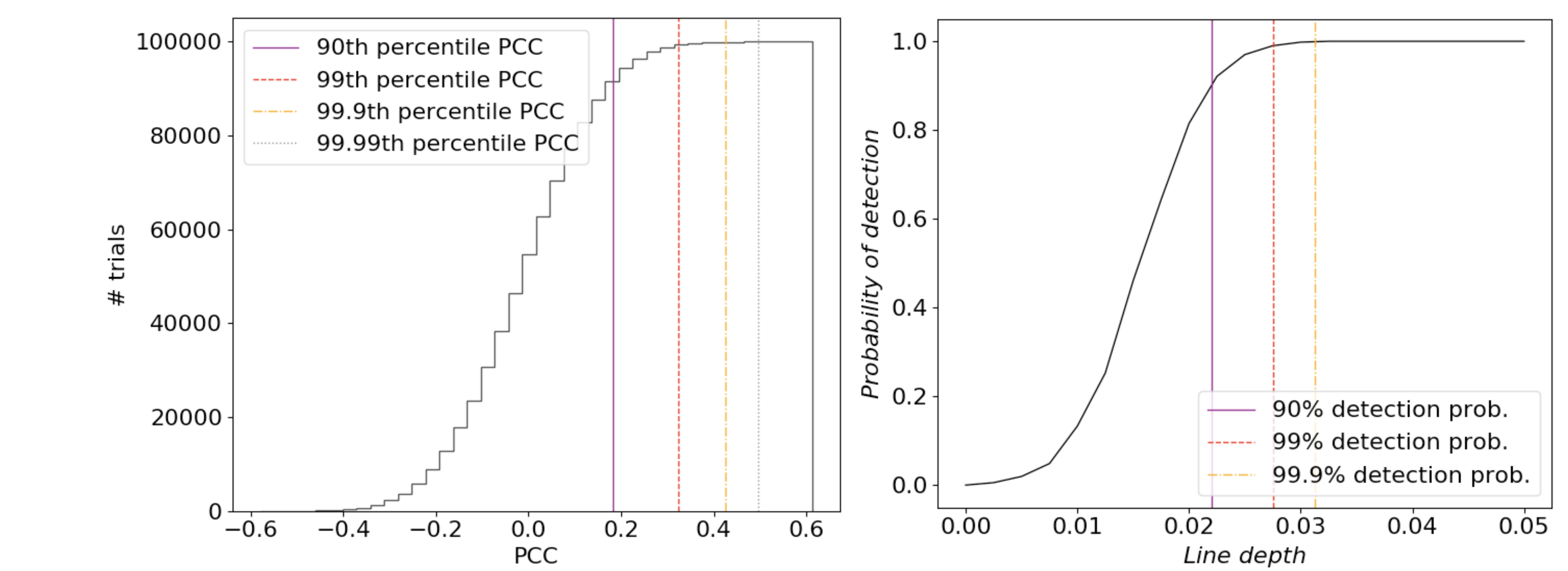}
  \caption{(left) A cumulative histogram for 100,000 trials to measure the PCC between a designated calibration pixel $\ln({I_\text{cal}})$ and pixels representing $\ln({I_\lambda})$ with only Gaussian noise scaled to $\sigma = 0.01$. (right) The probability of detecting a signal as a function of telluric line depth in the presence of the same level of Gaussian noise. The purple solid, red dashed, orange dash-dot and grey dotted lines show the 90, 99, 99.9, and 99.99 limits respectively.}
  \label{fig:k-test}
\end{figure}

The selection process begins by profiling the false positive rates of different values of $k$. The correlation  between a calibration pixel and a noise pixel in the data set is simulated by generating $n = 51$ points of the form [$\ln(I_{cal})$, $\ln(I_\lambda)$]. The values of $\ln(I_{cal})$ evenly fill the range $[-1, 0]$ and represent a range of possible calibration line depths, while values of $\ln(I_\lambda)$ are drawn at random from a Gaussian distribution with $\sigma = 0.01$, representing shot noise typical of the CHIRON spectra (\snr$\sim 100$). The PCC for each set is recorded, and the process repeated for 100,000 trials. The results are summarized in Figure~\ref{fig:k-test}. For the level of simulated noise, roughly 1\% of pixels yield a PCC of $0.323$; 0.1\% of pixels have a PCC above $0.425$ and fewer than $0.01$\% of pixels generate a PCC $> 0.506$. Since single and double pixel clusters with PCC above the threshold are rejected, assuming that each pixel's noise is independent, a threshold of $k = 0.425$ has just a $0.1\%^3 = 10^{-7}\%$ change of generating a false positive. Since the CHIRON spectrum has about 200,000 pixels, this threshold has just a 0.02\% chance of generating a false positive.

Once a threshold PCC is established, the minimum line depth detectable under the threshold in spectra with \snr\ $\sim 100$ is evaluated. A PCC threshold that is too high will fail to detect shallow lines (generating false negatives), reducing the sensitivity of the model. We again generated points representing pixels from 51 spectra with the form [$\ln(I_{cal}), \ln(I_\lambda)$]. The calibration line depth, $\ln(I_{cal})$, was again evenly distributed across the range $[-1, 0]$, while the pixels representing $\ln(I_\lambda)$ were scaled according to $\ln(I_\lambda) = c \cdot \ln(I_{cal})$. By randomly selecting values of $c \in [0, 0.07]$, these points represent telluric line depths $\le 7$\%. Gaussian noise consistent with \snr\ $\sim 100$ was then added to $\ln({I_\lambda})$, and the percentage of time that the PCC was greater than $k$ for pixel pairs was recorded. This simulation was repeated for 100,000 trials, and the results show that 90\% of lines deeper than 2.3\% and 99.9\% of lines 3\% of the continuum will be identified with the linear regression method described here (Fig~\ref{fig:k-test}, right). However, there is a precipitous drop in our ability to model tellurics with line depths shallower than 2\%. This result is, of course, dependent on the \snr\ of the training population and should improve if the training set had higher \snr\ and better continuum normalization. 

\subsubsection{SELENITE's Water Telluric Model}
The steps taken to identify and model water tellurics {in Section~\ref{ref:water_telluric_theory}} are summarized below.

\begin{enumerate}
    \item The PCC of each pixel's growth with a calibration pixel is calculated. A threshold PCC, $k$, is established, and pixels with PCC $> k$ are flagged as significant.
    
    \item Single or double pixels with PCC $> k$ are rejected as spurious.
    
    \item The training data set is coadded and a peak detection algorithm is applied to each cluster of more than three pixels. Clusters which do not contain a peak are rejected as tellurics.
    
    \item Any cluster of flagged pixels with no other cluster with 10\angstrom{} is rejected as a telluric feature.
    
    \item Linear regression is carried out on pixels that are flagged as tellurics to measure $m^\lambda_\text{cal}$ relative to a pre-identified calibration pixel. The wavelength, regression coefficient, PCC and water/non-water classification of each flagged pixel is then stored in a database.
\end{enumerate}

{The wavelength, linear coefficient, PCC, and a flag identifying the pixel as water is stored for each pixel that has passed the selection criteria for water tellurics is stored in a database. Table~\ref{tab:water_database} lists an excerpt of a database generated from the training data's content using the 5901.6\angstrom{} telluric as a calibrator.}  To generate a model of telluric water lines, the intensity of the central pixel in a calibration line is measured and information in the database is used to generate water tellurics for every pixel in the spectrum: 

\begin{align}
\ln{I_{\lambda_i}} = 
\begin{cases}
m^{\lambda_i}_\text{cal} \cdot \ln{I_\text{cal}} &when\ \lambda_i \in valid\ peak \wedge PCC_\lambda > k \\
0 &otherwise
\end{cases}
\end{align}

\noindent
where $m^{\lambda_i}_\text{cal}$ is the ratio of effective cross-section for absorption at $\lambda_i$ relative to $\lambda_\text{cal}$ is the effective cross-section of the calibration line wavelength (or the weighted average for an ensemble of calibration lines), $I_\text{cal}$ is the intensity at the calibration line wavelength,  and $k$ is the threshold correlation coefficient indicating telluric presence. {Generation of the telluric water model takes less than 3 minutes on a 2015 Macbook Air with a 2.2 GHz Intel Core i7 processor and 8GB of 1600 MHz DDR3 RAM} and allows for identification of variable numbers of telluric-contaminated pixels, depending on the PWV. 

{This is valuable since, as Figure~\ref{fig:o29-o44-growth} shows, water telluric size can vary by an order of magnitude. On night with high PWV, at a threshold $k$ of 0.425 (see Section~\ref{sec:bootstrap}), up to $\sim 4150$ pixels in our training spectra were contaminated, $3.1\%$ of pixels under 6800\angstrom{}. On dry nights, as few as $\sim 1700$ pixels were contaminated, $1.2\%$ of pixels under 6800\angstrom{}. This is a savings of $\sim 75\%$ of an order.}

\begin{table}[ht]
    \centering
    \vspace{-2mm}
    \begin{tabular}{c c c c}
        $\lambda_i$ &$\sigma_\lambda /  \sigma_{5901.6\angstrom{}}$ & PCC & Species Flag\\
        \hline
        5898.12061  &  0.49523  &  0.992 & W \\
        5898.14209  &  0.89206  &  0.994 & W \\
        5898.18457  &  0.66062  &  0.991 & W \\
        5898.20556  &  0.34039  &  0.937 & W \\
        5898.99121  &  0.47828  &  0.977 & W \\
        \hline
    \end{tabular}
    \caption{An excerpt from the telluric database generated for our training spectra.}
    \label{tab:water_database}
    \vspace{-8mm}
\end{table}

%\begin{deluxetable}{lclc}[!tbhp]
%    \tablecaption{\label{tab:Tbl1}}
%    \tablehead{ 
%    \colhead{$\lambda_i$}  &  %\colhead{$\sigma_\lambda /  \sigma_{5901.6\angstrom{}}$}  &  \colhead{PCC} & Species
%    }
%    \startdata
%    
%    \enddata
%
%\end{deluxetable}

\subsubsection{Identifying and Modelling Water Microtellurics}
\begin{figure}[!tbhp]
    \centering
  \includegraphics[clip, trim=0in 7.25in 2in 0in, width=7in]{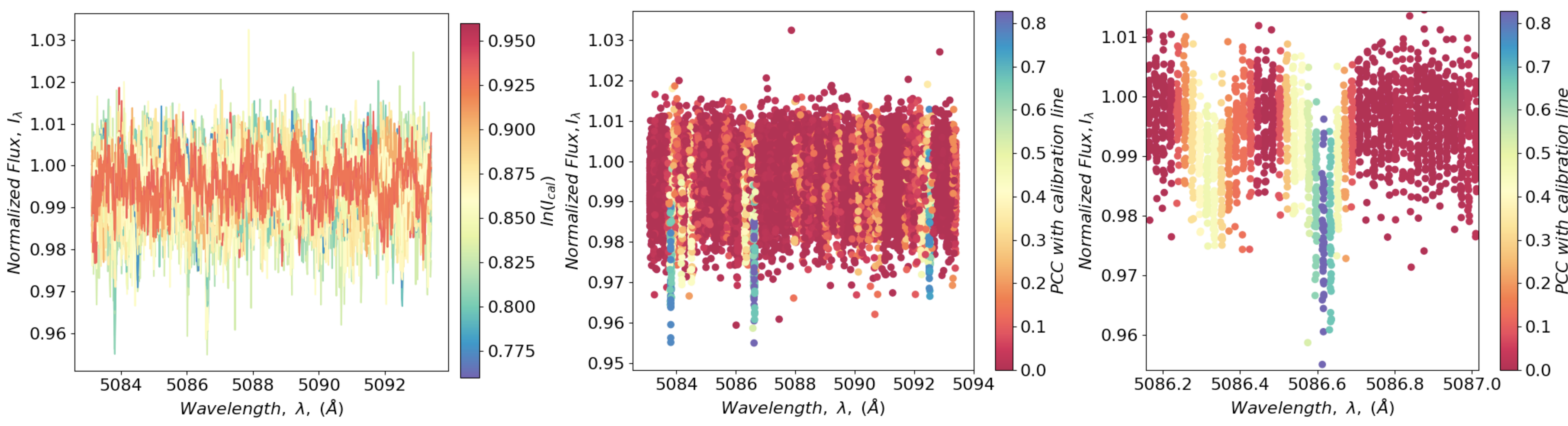}
  \caption{(left) Segments of 51 overplotted CHIRON spectrum in the wavelength range between 5082 - 5094\angstrom{}, color-coded by airmass. It is difficult to pick out telluric lines in this image. However, when the pixels are color-coded according to the PCC (middle), several microtellurics can be detected with high confidence. Zooming in on the wavelength segment at 5086\angstrom{} (right), the correlated pixel structure for identified weak telluric lines appears to be cleanly identified.}
  \label{fig:spectra-o14-and-corr}
\end{figure}

{SELENITE} is successful at identifying relatively shallow telluric features.
Figure~\ref{fig:spectra-o14-and-corr} shows the training set spectra for the wavelength range between $5075\angstrom{}$ and $5120\angstrom{}$. From the NSO atlas (Figure~\ref{fig:nso_microtell}) it is clear that this wavelength range should only contain weak microtelluric lines. Spectra in Figure~\ref{fig:spectra-o14-and-corr} (left) are color-coded by the intensity of the calibrating water telluric line at 5898.16\angstrom{} and it is difficult to see correlated growth for any microtelluric lines. However, when the pixels in each spectrum are color-coded by the strength of the PCC (regressed against a pixel in the core of the 5898.16\angstrom{} line), even telluric lines with a depth close to the photon noise in the continuum emerge with high confidence (Figure~\ref{fig:spectra-o14-and-corr}, middle). A close-up view (see right panel of Figure~\ref{fig:spectra-o14-and-corr}) highlights a detected microtelluric line with a depth only slightly greater than the photon noise.

\begin{figure}[!tbhp]
    \centering  
  \includegraphics[clip, trim=0in 7.25in 10.25in 0in, width=7in, ]{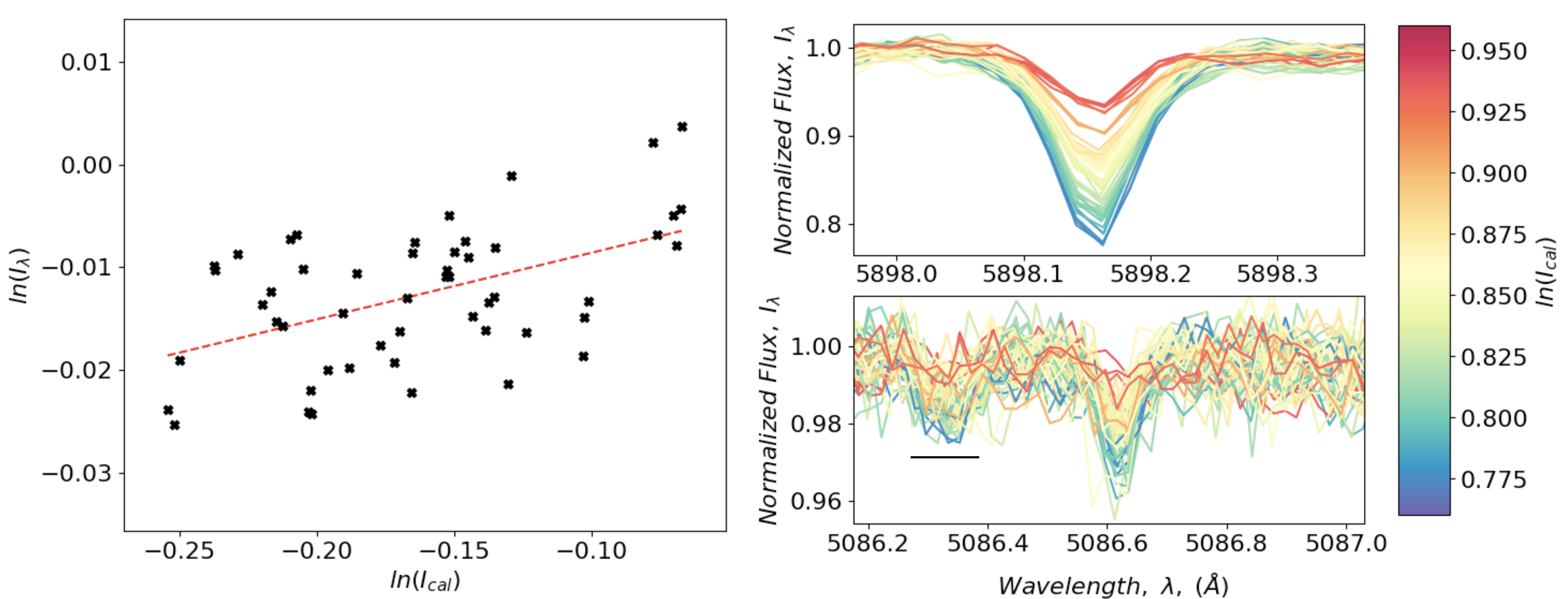}
  \caption{(left) The correlation between a microtelluric line intensity at 5086.3\angstrom{} and a telluric calibration line at 5898.16\angstrom. (right-top) A plot of a set of telluric lines at 5898.16\angstrom{} for our 51 Bstar spectra, color-coded by line depth. (right-bottom) The microtelluric line at 5086.3\angstrom{} (underlined). As before, the spectra for the microtelluric line are color-coded by the intensity of the calibrating telluric line, emphasizing the correlation.}
  \label{fig:calpx-o13-growth}
\end{figure}

{Moreover, SELENITE is accurate for microtellurics, whose depth is close to the shot noise of the spectra.} As an example, the pixel intensity at the center of a shallow microtelluric line is plotted against the pixel intensity of the 5898.16\angstrom{} calibration line in Figure~\ref{fig:calpx-o13-growth}. Following the format for Figure~\ref{fig:o29-o44-growth}, the telluric spectra in the wavelength region around 5898.16\angstrom{} and the spectra near 5086.3\angstrom{} (Figure~\ref{fig:calpx-o13-growth} right) are color-coded according to the depth of the 5898.16\angstrom{} line. 
{The linear regression between the calibration line and the underscored microtelluric line at 5086.3\angstrom{} is shown in the left panel of Figure~\ref{fig:calpx-o13-growth} and models the intensity of the microtelluric line with a mean SSE of 0.009, comparable to the \snr{} of the spectrum.}

\subsubsection{Using an Arbitrary Pixel as a Calibrator}
A powerful feature of SELENITE is that any arbitrary pixel or ensemble of pixels in the database can be substituted for the calibration pixel without requiring additional analysis by dividing each linear coefficient by the scale factor from the original calibration pixel to the new calibration pixel. As an example, Equation~\ref{eqn:model-trns} shows a how model based on calibration $A$ can be transformed to a model based on calibration line $B$. 

\begin{equation}
\label{eqn:model-trns}
\ln{I_{\lambda_i}} = \frac{m^{\lambda_i}_\text{A}}{m^\text{B}_\text{A}} 
                 \cdot \ln{I_\text{B}}
\end{equation}

The linear coefficients in the regression model were derived with B-stars (telluric stars) because these spectra have both high \snr\ and few spectral lines. However, once the linear coefficients have been derived, the coefficients can be used to model telluric contamination in spectra of later type stars as long as the selected telluric calibration line is isolated from the stellar absorption lines or the stellar absorption feature is well enough known (for example, by spectral synthesis modeling) that it can be divided out. The ability to use the database to switch between different calibrating pixels (described above) offers critical flexibility for modeling tellurics in spectra of late type stars. 

\subsection{Results for Water Tellurics}
\label{subsec:water_results}
%{We now evaluate the water telluric model. We measure: \textbf{(a)} its goodness of fit with the reduced chi square test statistic (Section~\ref{subsubsec:goodness_of_fit}), \textbf{(b)} (Section~\ref{subsubsec:microtelluric_performance}), and finally discuss \textbf{(c)} the relative contribution of PWV and airmass to water telluric line depth (Section~\ref{subsubsec:PWV_v_z}).}

\subsubsection{Model Goodness of Fit}
\label{subsubsec:goodness_of_fit}
{We evaluate SELENITE's goodness of fit using the B star HR3982's telluric spectrum. The HR3982 spectrum used was generated by averaging 3 unique observations taken over 40 min to drive} {up} {its S/N. Goodness of fit was measured using the reduced chi squared ($\chi_{red}^2$) test statistic. HR3892's observed flux was treated as the true model, $F_{obs, i}$, SELENITE's model of the flux as the "data", $F_{model, i}$ and the error calculated by the data reduction pipeline (0.75\% of continuum), scaled by \textbf{(a)} the root of the number of spectra coadded ($\sqrt{3}$) and \textbf{(b)} the root of model's flux ( $\sqrt{F_{model, i}}$)  as the statistical errors, $\sigma_{model, i} = 0.0075 / \sqrt{3 F_{model, i}}$.}

{First, to estimate the data quality independent of telluric removal, we measured the $\chi_{red}^2$ of a 3200px wavelength range unaffected by telluric lines, 4892\angstrom{}-4952\angstrom{}, with unity. We found a $\chi_{red}^2$ of 1.03, suggesting that our errors were well-calibrated. Next, the  $\chi_{red}^2$ of our model's fit in a 3200px wavelength range with heavy water tellurics, 6472\angstrom{}-6545\angstrom{} was measured. This range was chosen because \textbf{(a)} it contains the most intense water tellurics bluewards of 6800\angstrom{} and \textbf{(b)} it was free from stellar features. Only pixels where a telluric was detected were included in the $\chi_{red}^2$ calculation. A 25px range from 6521.5\angstrom{}-6522.5\angstrom{} was found to have errors $20\times$ higher than any other error, this region was flagged as an outlier and excluded. The $\chi_{red}^2$ of the telluric model was found to be 1.25. In particular, the line cores were fit well, with a $\chi_{red}^2$ of 1.11.  To reach a similar $\chi_{red}^2$ in the affected and unaffected region, errors in the affected region need to be increased by $\sim 10.5\%$.}

\begin{figure}[!tbhp]
  \centering
  \includegraphics[clip, trim=0in 7.5in 13in 0in, width=6.5in]{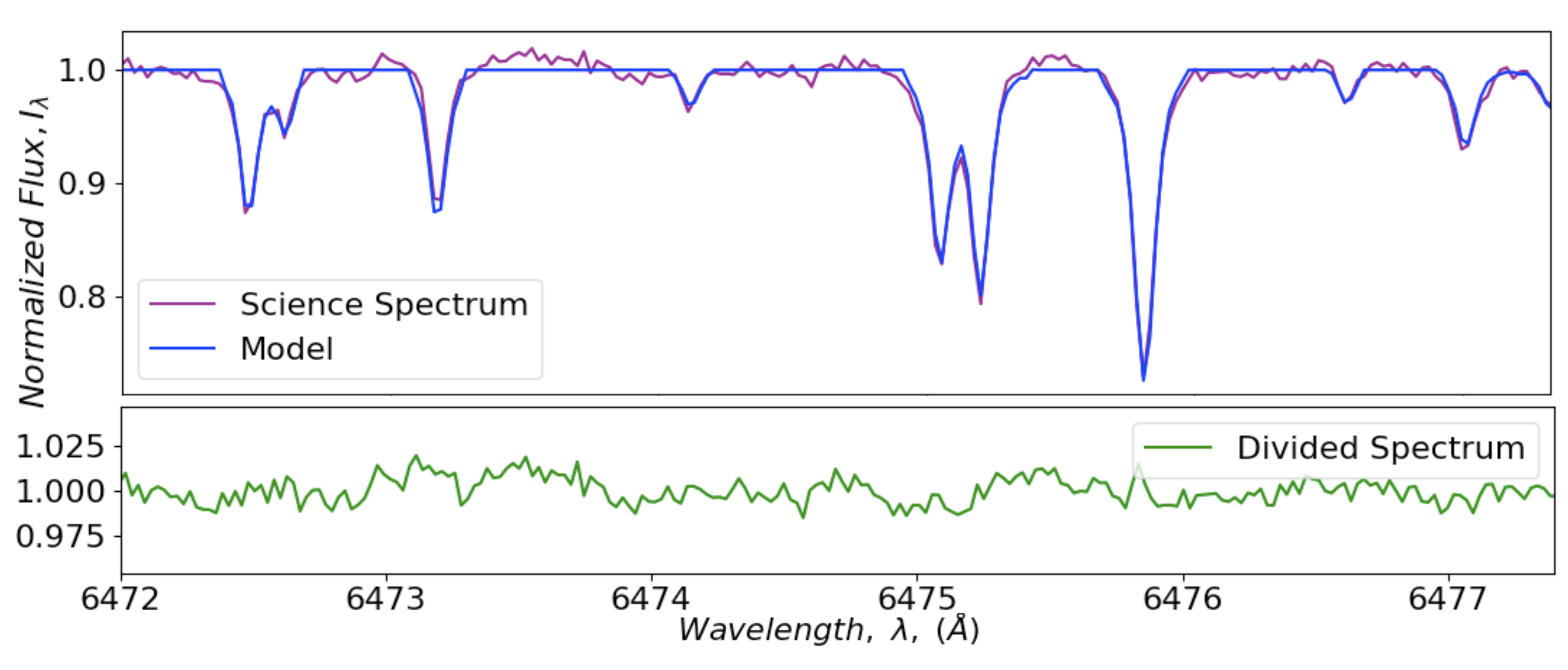}
  \vspace{-2mm}
  \caption{{(top) Excerpt of SELENITE's fit (blue) to HR3982's science spectrum, (purple). The fit's goodness is quantified in Section~\ref{subsubsec:goodness_of_fit}. (bottom) Residuals generated by dividing out the telluric model. The residuals deviate from unity by $\sim 1\%$ of the continuum on average, comparable to unaffected regions of the spectrum.}}
  %\caption{(top) The CHIRON spectrum for five co-added B-star with telluric contamination is plotted as a  purple line. A single input parameter, the depth of the central pixel at the core of the underlined line at 5898.16\angstrom{}, was measured and used to generate a telluric model for every pixel in this spectrum using the method described in this paper. (bottom) the residuals between the model and the spectrum with a standard deviation of 0.0053, comparable to the \snr\ of the coadded B-star spectrum.}
  \label{fig:w-modelling-chiron}
  \vspace{-4mm}
\end{figure}

{Figure~\ref{fig:w-modelling-chiron} (top) plots a 5\angstrom{} excerpt from the affected region, with HR3982's spectrum shown in purple and our model shown in blue. The fit's residuals deviate from unity by $1.0\%$ on average, comparable to the unaffected regions of the spectrum and the performance of radiative transfer codes. \citep{UlmerMoll2018}. One potential flaw in our model is that modelling all points without significant telluric signal as unity creates discontinuities in the telluric wings, however, Figure~\ref{fig:w-modelling-chiron} (bottom) indicates these discontinuities are small, and most users will prefer to mask affected pixels rather than dividing out.}

\subsubsection{Relative Contribution of PWV and Airmass to Water Line Depth}
\label{subsubsec:PWV_v_z}

\begin{figure}[!tbhp]
  \centering
\includegraphics[clip, trim=0in 7.5in 13.5in 0in, width=6in]{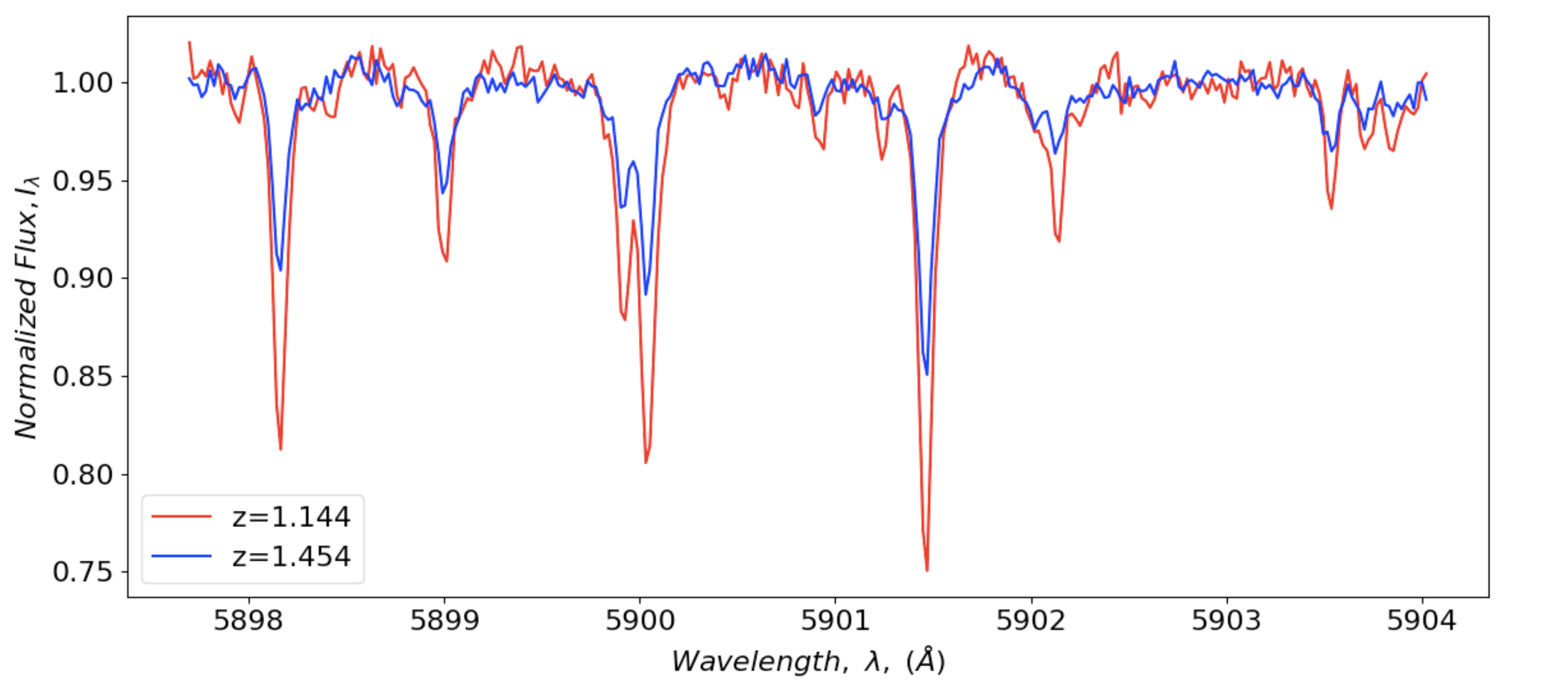}
  \caption{{A spectrum observed at an airmass of 1.144 (red) displays significantly deeper telluric lines than a spectrum observed at an airmass of 1.454 because of changes in PWV between observations.}}
  \label{fig:air-hum-contrast}
\end{figure}

{A further result is that the contribution PWV to water line depth generally dominates over airmass. As an example, Figure~\ref{fig:air-hum-contrast} shows that a low airmass (z=1.144) observation of the 5900\angstrom{} water lines can exhibit significantly greater line depth than a subsequent higher airmass (z=1.454) observation because of changes in PWV. While the water column density for an observation depends on both the average number density of absorbers along the line of sight (PWV) and the path length (airmass), PWV can vary by as much as an order of magnitude while airmass generally ranges between 1 and 2. In general, water line depth only weakly correlates with airmass. This lack of correlation can be exploited to distinguish water and non-water lines.}

%An interesting attribute of water line growth is the relative importance of airmass and PWV. According to Equation~\ref{eqn:lnI}, the water column density for an observation depends on both the average number density of absorbers along the line of sight (i.e. PWV) and the path length (i.e.,  airmass). However, variability in PWV has a stronger impact on telluric line formation than the changing path length due to airmass. This is illustrated in Figure~\ref{fig:air-hum-contrast}, where water telluric lines are plotted for two different nights. The telluric lines observed at a lower airmass (z=1.144) are significantly deeper than the same water tellurics observed at higher airmass (z=1.454), demonstrating the strong impact of PWV. The weak correlation between water telluric lines and airmass is helpful in distinguishing water and non-water lines.

\subsection{Non-water Tellurics} \label{subsec:non_water_tellurics}
In this section, telluric absorption lines from molecules other than water are considered. Like water tellurics, each non-water telluric can be modeled by the radiative transfer equation for a plane parallel atmosphere and thus its signal intensity given by $\sigma_{\lambda_i} \cdot n_j \cdot z$, where $n_j$ is the number density of the molecular species, $j$.

\begin{figure}[!tbhp]
  \centering
  \includegraphics[clip, trim=0in 7.25in 10.5in 0in, width=7in]{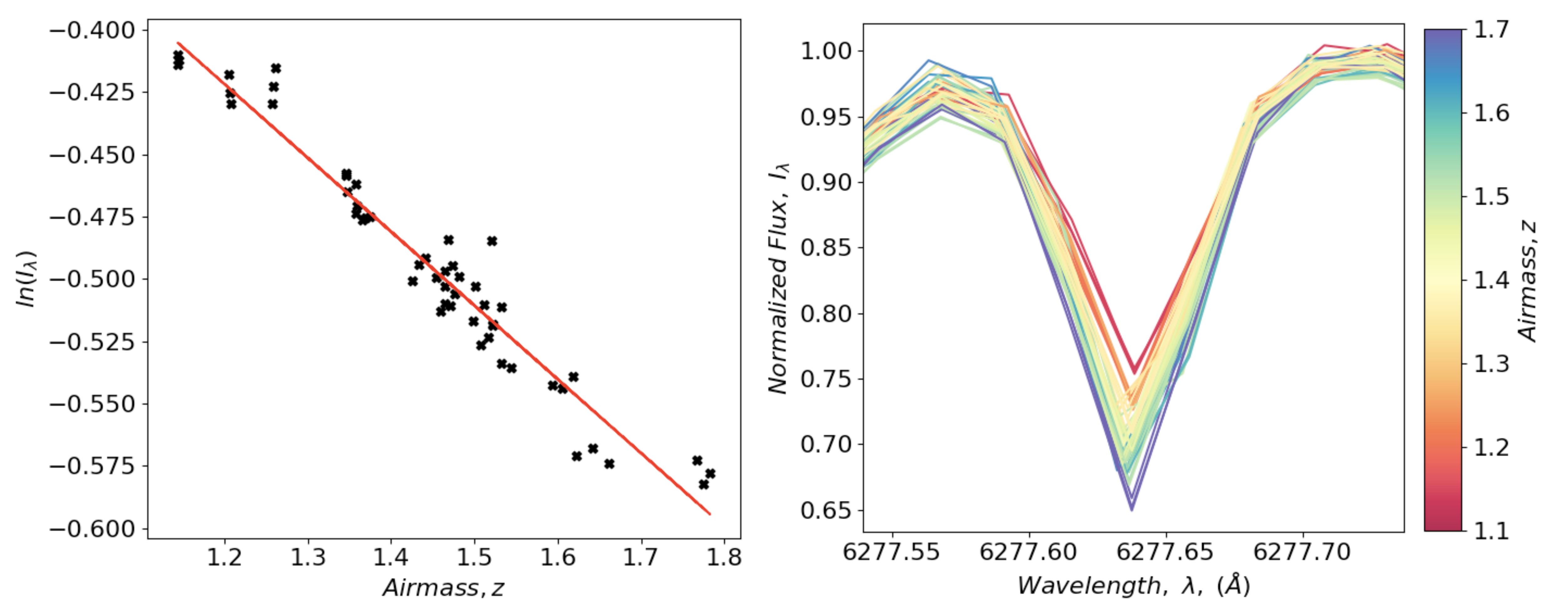}
  \caption{(left) The correlation between the 6277.6\angstrom{} oxygen feature and airmass. (right) The telluric oxygen feature at 6277.6\angstrom{} for our set of 51 CHIRON spectra, color-coded by airmass. }
  \label{fig:o51-airmass-plot}
\end{figure}

Unlike water tellurics, however, non-water tellurics have no equivalent of PWV. Ignoring small seasonal variations in gases such as ${\rm CO_2}$, $n_j$ is spatially and temporally fixed. Each non-water species in the atmosphere is evenly distributed with a constant number density. Therefore the column density of non-water lines only varies with airmass: {\textbf{by measuring airmass, we can predict the depth of every non-water line in the spectrum.} As an example, Figure~\ref{fig:o51-airmass-plot} (right) shows that over our observed range of airmass ($z$ between $1.1 - 1.8$) the signal intensity of the oxygen telluric feature at 6277.7\angstrom{} (Figure~\ref{fig:o51-airmass-plot}, left) is well fitted by the linear regression model $\ln(I_{6277.7\angstrom{}}) = m \cdot z + b$. The slope of the regression model, $m$, measures $\sigma_{\lambda_i} \cdot n_j$. Another difference from the model for water lines is that the y-intercept (a fictitious extrapolation to zero airmass) is small, but non-zero.}  

{Like water lines, non-water lines can be identified by measuring the correlation of their growth with airmass. Each pixel whose growth's PCC with airmass is above a threshold, $k$, is assumed to have non-water telluric and undergoes the same procedure as water telluric pixels. Again, this potentially allows for the detection of tellurics not listed in the HITRAN database.} 

{Non-water lines can be readily distinguished from water lines because non-water lines have a low correlation with the water calibration pixels but a high correlation with airmass, and vice versa for water lines (see Section~\ref{subsubsec:PWV_v_z}). Separating components that vary with airmass from those that don't is a benefit of SELENITE that might well be useful outside the scope of this paper, which as in the near IR, where $\text{H}_2\text{O}$, $\text{CO}_2$ and $\text{CH}_4$ lines mix. When a water and non water line blend, the composite line can have a significant correlation with both the water calibrator and airmass. A regression model is not fit to composite lines, but they are flagged in the database.}

\subsection{Results for Non-Water Tellurics}
\label{subsec:non_water_results}
\begin{figure}[!tbhp]
  \centering
  \includegraphics[clip, trim=0in 9in 11in 0in, width=\linewidth]{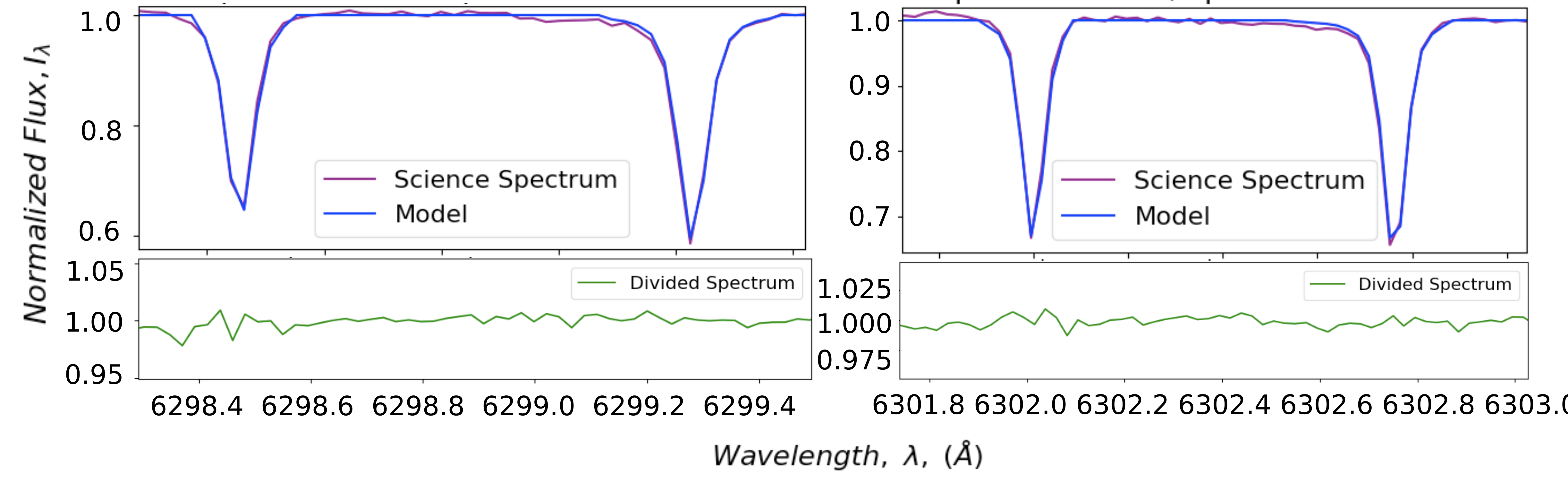}
  
  \caption{{(top) Excerpt of the model's fit (blue) to two oxygen doublets in the $\text{O}_2\ \gamma$ atmospheric band in HR3982's science spectrum (purple). The fits goodness is quantified in Section~\ref{subsec:non_water_results} (bottom) Residuals generated by dividing out the telluric model. The residuals deviate from unit by $\sim 0.75\%$ of the continuum on average, comparable to unaffected regions of the spectrum.}}

  %\caption{Three CHIRON B star spectra that were not included in the training set were coadded to increase \snr{} and plotted in purple. The regression model for this part of the spectrum accurately reproduces the strength of the oxygen lines from $6298- 6306$\angstrom.}
  \label{fig:non-water-model-vs-actual}
\end{figure}

{We evaluate SELENITES's gooodness of fit by using the B star HR 3982's telluric spectrum following the procedure described in Section~\ref{subsubsec:goodness_of_fit}. This time, however, we measured the $\chi_{red}^2$ of the models fit from 6257\angstrom{}-6328\angstrom{}, a 3200px wavelength range which encompasses the heart of the 6280\angstrom{} $\text{O}_2$ $\gamma$ atmospheric band. Only pixels where a non-water telluric was detected were measured. The  $\chi_{red}^2$ of the telluric model was found to be 1.17. To reach a similar $\chi_{red}^2$ in the affected and unaffected region, errors in the affected region need to be increased by $2.0\%$. Figure~\ref{fig:non-water-model-vs-actual} plots the model's fit to two oxygen doublets in HR3982's $\text{O}_2\ \gamma$ atmospheric band. The fit's residuals deviate from unity by about $\sim 0.75\%$ on average, comparable to unaffected regions of the spectrum.}

  {Unfortunately, there are no non-water species with telluric lines other than oxygen bluewards of 6800\angstrom{}, so we cannot evaluate our model on other species. Fundamentally, however, any well mixed non- water species should in theory behave as oxygen does.}

\subsection{Modelling Tellurics in a K Dwarf Spectrum}
\label{subsec:K_dwarf_modelling}

{Late-type stars display complex absorption features. These absorption features do not complicate SELENITE's non-water modelling, which only measures airmass, but they do complicate water modelling, since they may blend with a calibration pixel's line. To compensate for the loss of any given calibration pixel, a large ($50+$) ensemble of potential calibration pixels are given in the database.}

{Calibration pixels which are blended with stellar lines are identified and removed as follows. Initially, a telluric model is built by regression against the average of all calibration pixel depths. If any calibration line is blended with a stellar line, the regression model with overestimate PWV and the depth of every non-blended water line, but will underestimate the depth of the blended calibration pixel's line. This calibration pixel can then be removed from the calibration set, and the process repeated until the calibration set stablizes. Empirically, we find that as long as just 25\% of calibration pixels remain, SELENITE generates a good fit.}

%{Late-type stars have complex absorption features, which complicate the modelling of telluric lines. This is not an issue for non-water lines, which grow with increasing air mass, but it is a challenge for generating a water telluric model, which requires a calibrating water line. To check for stellar line blending, a set of prospective calibration lines with regression coefficients in the telluric database is tested. Calibrating telluric lines that produce a telluric model that is too deep leave large residual noise between the model and the observation and are likely blended with a stellar absorption line. Each of the prospective water telluric lines in a set is tested by generating a telluric model and calculating the residual noise for pixels with telluric contamination. Calibrating lines that produce large residuals are rejected from the original set and calibrating lines that minimize residual noise are retained. Empirically, we find that water tellurics imprinted in cool stellar spectra can be modelled with the linear regression method described here, provided that some of the calibration lines are unblended.}

{We evaluate SELENITE's fit on late-type stars with the K-dwarf $\alpha$ Centauri B. We measured the $\chi_{red}^2$ of the models fit at the 6450\angstrom{} water band described in Section~\ref{subsubsec:goodness_of_fit}. This measurement, however, was complicated by $\alpha$ Centauri B's stellar lines: if a telluric line is blended with a stellar line, the model's fit will appear incorrect. This problem was overcome by noticing that changes in the Earth's barycentric velocity will substantially shift the stellar lines in two observations of $\alpha$ Centauri B taken months apart while leaving the telluric lines in the same position. Tellurics that are blended in the first observation will often be unblended in the second observation, and vice versa.} 

\begin{figure}[ht]
  \centering
  \includegraphics[clip, trim=0in 5.5in 11.5in 0in, width=\linewidth]{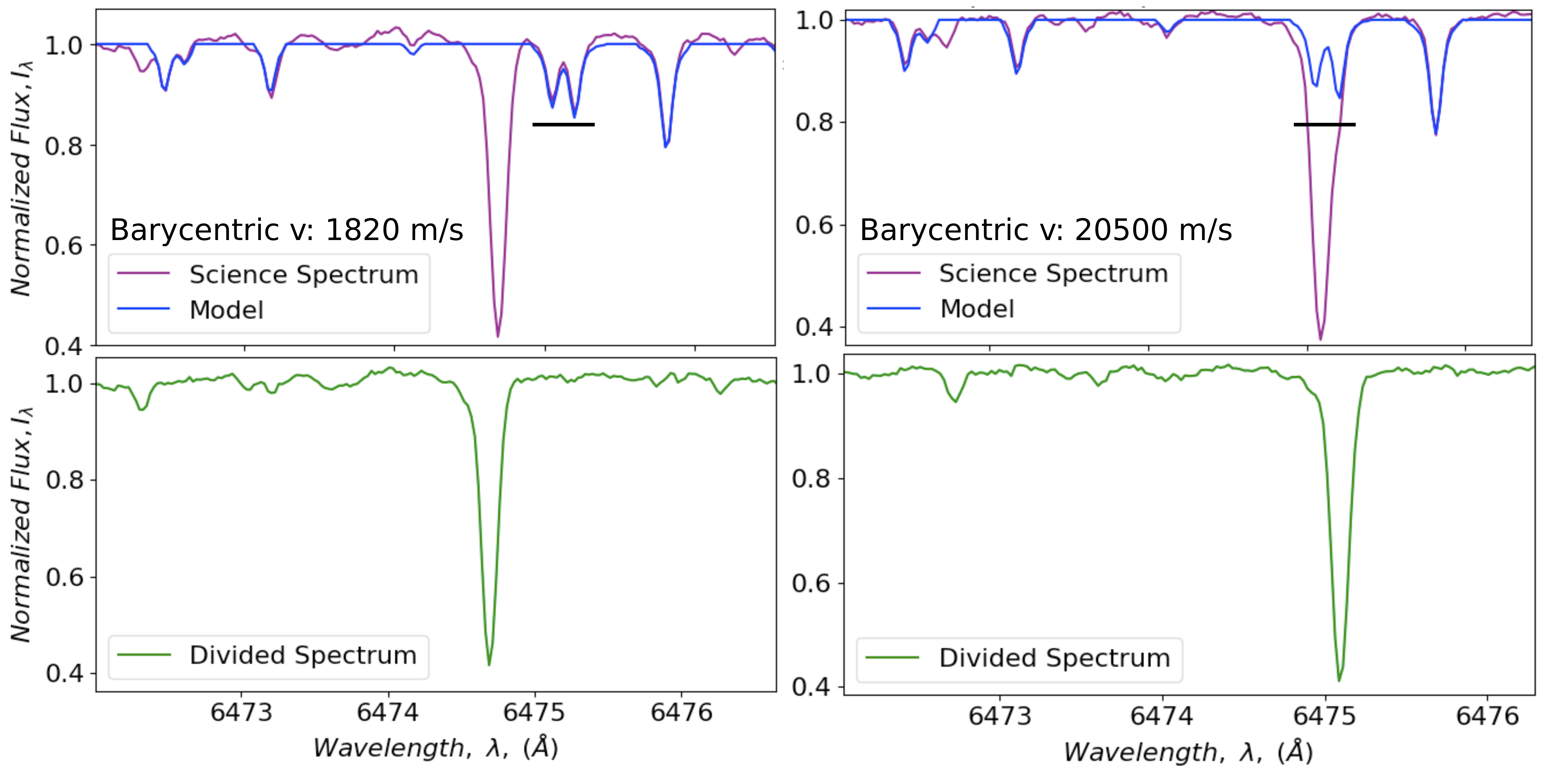}
  \caption{{ (top) Excerpt of the model's fit (blue) to observations of $\alpha$ Centauri B science spectrum (purple) at barycentric velocities of 1820 m/s and 20500 m/s. The apparent misfit underlined in the 20500 m/s spectrum is revealed to be a good fit in the 1820 m/s spectrum when the deep stellar line at 6475\angstrom{} shifts bluewards (bottom) The model's residuals  reveal that the two underlying spectra are the same.}}
  \label{fig:cool-stellar-fitting}
\end{figure}

{To illustrate, Figure~\ref{fig:cool-stellar-fitting} (top) shows SELENITE's fit to two observations of $\alpha$ Centauri B, at barycentric velocities of 1860 m/s and 20500 m/s, for the same 5\angstrom{} wavelength range shown in Section~\ref{subsubsec:goodness_of_fit}. In the 20500 m/s observation, the deep line at 6475\angstrom{} seems ill fit by the model's pair of water lines (underlined), but in the 1860 m/s observation the deep line has shifted, revealing that it was a stellar line blended with a pair of water line which the model now fits well. The fit's residuals, shown in Figure~\ref{fig:cool-stellar-fitting} bottom, show that when tellurics are removed the two spectra are indeed the same. Where the residuals do not contain a stellar line, they deviate from unity by an average of 1.1\%, comparable to the results of a radiative transfer code.}

{When we compute $\chi_{red}^2$, if the spectrum grossly deviates from a pixel fit (by $3.0\%$ or more of the continuum) we assume that the pixel is blended with a stellar line and reject it. Following this procedure, we found an $\chi_{red}^2$ of 2.95 and 3.17 for the 1840 m/s and 20500 m/s $\alpha$ Centauri B observations. This fit, while acceptable, is somewhat poorer than HR3982's fit, in large part because telluric lines often blend with stellar line tails, disrupting their profile slightly. For example, the wings of the small telluric at 6472.5\angstrom{} (at the far left of Figure~\ref{fig:cool-stellar-fitting}) are blended with a small stellar telluric, inflating the measurement of $\chi_{red}^2$.}

%{As an example, Figure~\ref{fig:cool-stellar-fitting} shows the telluric model fit to the 6480\angstrom{} segment of spectrum for the K0 star, $\alpha$ Centauri B. Prospective calibration lines that produced 4\% residuals were found to be blended with stellar lines and were rejected. After eliminating the blended prospective calibration lines as described above, the average residual for pixels with modeled tellurics is 1.0\% of the continuum.}

\section{Discussion} 
{Because of the barycentric velocity of the Earth, telluric lines raster across the stellar line profiles in time-series Doppler measurements. Even shallow microtelluric features will degrade the fidelity of high-resolution spectra and may contribute up to 0.5 \ms\ to the RV error budget. Since the Earth induces a radial velocity of 10 \cms\ in the Sun, telluric contamination is a significant challenge in the search for analogs of our world. In this paper, we present SELENITE, an empirical technique for identifying and modelling telluric features in the optical (4500\angstrom{}-6800\angstrom{}), using the observations: \textbf{(a)} water tellurics grow proportionally to PWV and therefore proportionally to each other and \textbf{(b)} non-water tellurics grow proportionally to airmass. Water tellurics are identified by looking for pixels whose growth correlates with a known \textit{calibration water telluric} and modelled by regression against it. Non-water tellurics are identified by looking for pixels whose growth correlates with airmass and modelled by regression against it. SELENITE has several advantages over the alternatives:}

%\deleted{Telluric absorption lines have depths that range from 1\% of the continuum in the optical bandpass to completely opaque lines in the near infrared}. Because of the barycentric velocity of the Earth, telluric lines raster across the stellar line profiles in time-series Doppler measurements. Even shallow microtelluric features will degrade the fidelity of high-resolution spectra and may contribute up to 0.5 \ms\ to the RV error budget. Since the Earth induces a radial velocity of 10 \cms\ in the Sun, telluric contamination is a significant challenge in the search for analogs of our world. 

%{Identified tellurics and their regression models are stored in a database which can be used to generate a telluric model for a science spectrum, given its airmass and calibration telluric depth. By manipulating the constants in the database, any set of water tellurics can act as calibration tellurics, allowing the method to be applied to late-type stars where the original calibration telluric may be blended with a stellar line. Building this database only requires a one-time observation of a few dozen B-stars across a few nights. }

\begin{itemize}[itemsep=2pt,parsep=2pt]
    \item {\textbf{Runtime:} Once the database is built ($<3 min$ on a standard PC) fitting a spectrum takes several seconds, permitting SELENITE to be used at the telescope to help guide observing runs.}
    \item {\textbf{Observing time:} Unlike standard stars, after a one time observation of a few dozen B stars to build the database, SELENITE requires no further observations, saving observing time.}
    \item {\textbf{Requires no atomic/molecular line data:} Unlike radiative transfer codes, SELENITE does not require atomic/molecular line data. This is useful because the literature suggests HITRAN is not always accurate. In particular, \cite{Seifahrt2010} notes: "Line data in HITRAN have strongly varying accuracy levels. Typical uncertainties of line positions range from a few to several hundred m/s, but can be as high as several km/s in extreme cases. Line strengths are rarely precise to the 1\% level." Further, \cite{Rudolf2016} find that inaccuracies in the HITRAN database frustrate their ability to model water lines accurately.}
    %\item {\textbf{Requires no external PWV measurements:} Many radiative transfer codes (e.g., TAPAS) require PWV measurement for best performance. Measuring PWV along the line of sight, however, is difficult: recent approaches include \cite{Baker2017}, who use a multi-band photometer to monitor atmospheric PWV and \cite{Li2018}, who monitor GPS signal strength to estimate PWV. By measuring PWV using the science spectrum's tellurics, SELENITE avoids costly and sometimes inaccurate PWV measurements.}
    \item {\textbf{Distinguishes tellurics that vary primarily with airmass from those that don't:} Although outside the paper's scope, this feature could be very useful in the near IR, where $\text{H}_2\text{O}$, $\text{CO}_2$ and $\text{CH}_4$ lines mix. }
\end{itemize}

{We acknowledge, however, that SELENITE has certain limitations. \textbf{First}, stellar features in the set of training B stars, (e.g., the Paschen and Brackett lines) will distort its model. This problem can be solved by interpolating over each absorption, at the cost of introducing additional uncertainity to regions of scientific interest. \textbf{Second}, SELENITE only varies with airmass and PWV. Other atmospheric phenomena which may affect line profiles (e.g., wind speed \citep{Caccin1985}) is not taken into account. Instrumental changes, such as a varying LSF, are also not considered, and can only be handled by rebuilding the database for each instrumental profile change. \textbf{Third}, SELENITE's PCC cutoff threshold produces discontinuities. While these discontinuities are small from CHIRON's high \snr{} data, at lower \snr{} a line's wings may not clear the PCC threshold, truncating them.}

{Despite these limitations, evaluations show that SELENITE provides excellent fits. The model's fit to regions of intense water tellurics and non-water tellurics in the B star HR3982 had $\chi^2_{red}$ of 1.25 and 1.17, and thus errors just 10.5\% and 2.0\% bigger than the continuum's fit to unity. Further, SELENITE's fits to the K-dwarf $\alpha$ Centauri B observations had $\chi^2_{red}$ of 2.95 and 3.17, despite the $\chi^2_{red}$ test statistic being inflated by stellar line blending, confirming that it provides a good fit to late-type stars. SELENITE's average residual is $1.0\%$ and $0.75\%$ for HR3982 and $1.1\%$ for $\alpha$ Centauri B, comparable to the residuals of radiative transfer codes \citep{UlmerMoll2018}.}

\section{acknowledgements}
Acknowledgements: The authors gratefully acknowledge enabling support from the following grants NSF-1616086, NSF-MRI0923441, NASA-NNH17ZDA001N-XRP, NASA-NNH11ZDA001N-OSS. NSO/Kitt Peak FTS data used here were produced by NSF/NOAO. 

%Simulation results have been provided by the Community Coordinated Modeling Center at Goddard Space Flight Center through their public Runs on Request system (http://ccmc.gsfc.nasa.gov). 

\facilities{CTIO: (CHIRON)}

\pagebreak
\bibliography{main.bib} 

\begin{thebibliography}{}
\expandafter\ifx\csname natexlab\endcsname\relax\def\natexlab#1{#1}\fi

\bibitem[{{Artigau} {et~al.}(2014){Artigau}, {Astudillo-Defru}, {Delfosse},
  {Bouchy}, {Bonfils}, {Lovis}, {Pepe}, {Moutou}, {Donati}, {Doyon}, \&
  {Malo}}]{Artigau2014}
{Artigau}, {\'E}., {Astudillo-Defru}, N., {Delfosse}, X., {et~al.} 2014, in
  \procspie, Vol. 9149, Observatory Operations: Strategies, Processes, and
  Systems V, 914905

\bibitem[{{Bender} {et~al.}(2012){Bender}, {Mahadevan}, {Deshpande}, {Wright},
  {Roy}, {Terrien}, {Sigurdsson}, {Ramsey}, {Schneider}, \&
  {Fleming}}]{Bender2012}
{Bender}, C.~F., {Mahadevan}, S., {Deshpande}, R., {et~al.} 2012, \apjl, 751,
  L31

\bibitem[{{Bertaux} {et~al.}(2014){Bertaux}, {Lallement}, {Ferron}, {Boonne},
  \& {Bodichon}}]{Bertaux2014}
{Bertaux}, J.~L., {Lallement}, R., {Ferron}, S., {Boonne}, C., \& {Bodichon},
  R. 2014, \aap, 564, A46

\bibitem[{{Blake} \& {Shaw}(2011)}]{BlakeShaw2011}
{Blake}, C.~H., \& {Shaw}, M.~M. 2011, \pasp, 123, 1302

\bibitem[{{Caccin} {et~al.}(1985){Caccin}, {Cavallini}, {Ceppatelli},
  {Righini}, \& {Sambuco}}]{Caccin1985}
{Caccin}, B., {Cavallini}, F., {Ceppatelli}, G., {Righini}, A., \& {Sambuco},
  A.~M. 1985, \aap, 149, 357

\bibitem[{{Cunha} {et~al.}(2014){Cunha}, {Santos}, {Figueira}, {Santerne},
  {Bertaux}, \& {Lovis}}]{Cunha2014}
{Cunha}, D., {Santos}, N.~C., {Figueira}, P., {et~al.} 2014, \aap, 568, A35

\bibitem[{{Figueira} {et~al.}(2010){Figueira}, {Pepe}, {Lovis}, \&
  {Mayor}}]{Figueira2010}
{Figueira}, P., {Pepe}, F., {Lovis}, C., \& {Mayor}, M. 2010, \aap, 515, A106

\bibitem[{{Fischer} {et~al.}(2016){Fischer}, {Anglada-Escude}, {Arriagada},
  {Baluev}, {Bean}, {Bouchy}, {Buchhave}, {Carroll}, {Chakraborty}, {Crepp},
  {Dawson}, {Diddams}, {Dumusque}, {Eastman}, {Endl}, {Figueira}, {Ford},
  {Foreman-Mackey}, {Fournier}, {F{\H u}r{\'e}sz}, {Gaudi}, {Gregory},
  {Grundahl}, {Hatzes}, {H{\'e}brard}, {Herrero}, {Hogg}, {Howard}, {Johnson},
  {Jorden}, {Jurgenson}, {Latham}, {Laughlin}, {Loredo}, {Lovis}, {Mahadevan},
  {McCracken}, {Pepe}, {Perez}, {Phillips}, {Plavchan}, {Prato}, {Quirrenbach},
  {Reiners}, {Robertson}, {Santos}, {Sawyer}, {Segransan}, {Sozzetti},
  {Steinmetz}, {Szentgyorgyi}, {Udry}, {Valenti}, {Wang}, {Wittenmyer}, \&
  {Wright}}]{Fischer2016}
{Fischer}, D.~A., {Anglada-Escude}, G., {Arriagada}, P., {et~al.} 2016, \pasp,
  128, 066001

\bibitem[{{Gullikson} {et~al.}(2014){Gullikson}, {Dodson-Robinson}, \&
  {Kraus}}]{Gullikson2014}
{Gullikson}, K., {Dodson-Robinson}, S., \& {Kraus}, A. 2014, \aj, 148, 53

\bibitem[{{Hadrava, P.}(2006)}]{Hadrava2006}
{Hadrava, P.} 2006, A\&A, 448, 1149

\bibitem[{{Jurgenson} {et~al.}(2016){Jurgenson}, {Fischer}, {McCracken},
  {Sawyer}, {Szymkowiak}, {Davis}, {Muller}, \& {Santoro}}]{Jurgenson2016}
{Jurgenson}, C., {Fischer}, D., {McCracken}, T., {et~al.} 2016, in \procspie,
  Vol. 9908, Ground-based and Airborne Instrumentation for Astronomy VI, 99086T

\bibitem[{{Maiolino} {et~al.}(1996){Maiolino}, {Rieke}, \&
  {Rieke}}]{Maiolino1996}
{Maiolino}, R., {Rieke}, G.~H., \& {Rieke}, M.~J. 1996, \aj, 111, 537

\bibitem[{{Pepe} {et~al.}(2013){Pepe}, {Cristiani}, {Rebolo}, {Santos},
  {Dekker}, {M{\'e}gevand}, {Zerbi}, {Cabral}, {Molaro}, {Di Marcantonio},
  {Abreu}, {Affolter}, {Aliverti}, {Allende Prieto}, {Amate}, {Avila},
  {Baldini}, {Bristow}, {Broeg}, {Cirami}, {Coelho}, {Conconi}, {Coretti},
  {Cupani}, {D'Odorico}, {De Caprio}, {Delabre}, {Dorn}, {Figueira}, {Fragoso},
  {Galeotta}, {Genolet}, {Gomes}, {Gonz{\'a}lez Hern{\'a}ndez}, {Hughes},
  {Iwert}, {Kerber}, {Landoni}, {Lizon}, {Lovis}, {Maire}, {Mannetta},
  {Martins}, {Monteiro}, {Oliveira}, {Poretti}, {Rasilla}, {Riva}, {Santana
  Tschudi}, {Santos}, {Sosnowska}, {Sousa}, {Span{\`o}}, {Tenegi}, {Toso},
  {Vanzella}, {Viel}, \& {Zapatero Osorio}}]{Pepe2013}
{Pepe}, F., {Cristiani}, S., {Rebolo}, R., {et~al.} 2013, The Messenger, 153, 6

\bibitem[{{Rothman} {et~al.}(2013){Rothman}, {Gordon}, {Babikov}, {Barbe},
  {Chris Benner}, {Bernath}, {Birk}, {Bizzocchi}, {Boudon}, {Brown},
  {Campargue}, {Chance}, {Cohen}, {Coudert}, {Devi}, {Drouin}, {Fayt}, {Flaud},
  {Gamache}, {Harrison}, {Hartmann}, {Hill}, {Hodges}, {Jacquemart}, {Jolly},
  {Lamouroux}, {Le Roy}, {Li}, {Long}, {Lyulin}, {Mackie}, {Massie},
  {Mikhailenko}, {M{\"u}ller}, {Naumenko}, {Nikitin}, {Orphal}, {Perevalov},
  {Perrin}, {Polovtseva}, {Richard}, {Smith}, {Starikova}, {Sung}, {Tashkun},
  {Tennyson}, {Toon}, {Tyuterev}, \& {Wagner}}]{Rothman2013}
{Rothman}, L.~S., {Gordon}, I.~E., {Babikov}, Y., {et~al.} 2013, \jqsrt, 130, 4

\bibitem[{{Rudolf, N.} {et~al.}(2016){Rudolf, N.}, {G\"unther, H. M.},
  {Schneider, P. C.}, \& {Schmitt, J. H. M. M.}}]{Rudolf2016}
{Rudolf, N.}, {G\"unther, H. M.}, {Schneider, P. C.}, \& {Schmitt, J. H. M. M.}
  2016, A\&A, 585, A113

\bibitem[{{Seifahrt, A.} {et~al.}(2010){Seifahrt, A.}, {K\"aufl, H. U.},
  {Z\"angl, G.}, {Bean, J. L.}, {Richter, M. J.}, \& {Siebenmorgen,
  R.}}]{Seifahrt2010}
{Seifahrt, A.}, {K\"aufl, H. U.}, {Z\"angl, G.}, {et~al.} 2010, A\&A, 524, A11

\bibitem[{{Smette} {et~al.}(2015){Smette}, {Sana}, {Noll}, {Horst}, {Kausch},
  {Kimeswenger}, {Barden}, {Szyszka}, {Jones}, {Gallenne}, {Vinther},
  {Ballester}, \& {Taylor}}]{Smette2015}
{Smette}, A., {Sana}, H., {Noll}, S., {et~al.} 2015, \aap, 576, A77

\bibitem[{{Tokovinin} {et~al.}(2013){Tokovinin}, {Fischer}, {Bonati},
  {Giguere}, {Moore}, {Schwab}, {Spronck}, \& {Szymkowiak}}]{Tokovinin2013}
{Tokovinin}, A., {Fischer}, D.~A., {Bonati}, M., {et~al.} 2013, \pasp, 125,
  1336

\bibitem[{{Ulmer-Moll, S.} {et~al.}(2019){Ulmer-Moll, S.}, {Figueira, P.},
  {Neal, J. J.}, {Santos, N. C.}, \& {Bonnefoy, M.}}]{UlmerMoll2018}
{Ulmer-Moll, S.}, {Figueira, P.}, {Neal, J. J.}, {Santos, N. C.}, \& {Bonnefoy,
  M.} 2019, A\&A, 621, A79

\bibitem[{{Vacca} {et~al.}(2003){Vacca}, {Cushing}, \& {Rayner}}]{Vacca2003}
{Vacca}, W.~D., {Cushing}, M.~C., \& {Rayner}, J.~T. 2003, \pasp, 115, 389

\bibitem[{{Vidal-Madjar} {et~al.}(1986){Vidal-Madjar}, {Ferlet}, {Gry}, \&
  {Lallement}}]{1986:Vidal}
{Vidal-Madjar}, A., {Ferlet}, R., {Gry}, C., \& {Lallement}, R. 1986, \aap,
  155, 407

\bibitem[{{Wallace} {et~al.}(1993){Wallace}, {Livingston}, \&
  {Hinkle}}]{Wallace1993}
{Wallace}, L., {Livingston}, W.~C., \& {Hinkle}, K. 1993, {An atlas of the
  solar photospheric spectrum in the region from 8900 to 13600 cm$^{-1}$ ($7350
  - 11230 {\AA}$) with decomposition into solar and atmospheric components and
  identifications of the main solar features}

\end{thebibliography}
%\listofchanges
\end{document}